\documentstyle[prb,aps,multicol]{revtex}

\begin{document}

\preprint{Preprint}
\draft{}

\title{Transition from Type-I to Type-II Superconducting Behaviour with
Temperature observed by $\mu$SR and SANS}

\author{C.~M.~Aegerter\cite{Adr}, H.~Keller \\}
\address{Physik-Institut der Universit\"at Z\"urich,
CH-8057 Z\"urich, Switzerland.\\ }
\author{S.~L.~Lee, C.~Ager, F.~Y.~Ogrin \\}
\address{School of Physics and Astronomy, University of St.~Andrews,
St.~Andrews, Fife KY169SS, UK.\\ }
\author{R.~Cubitt \\}
\address{Institut Laue-Langevin, Grenoble, France. \\}
\author{E.~M.~Forgan, W.~J.~Nuttall, P.~G.~Kealey, S.~H.~Lloyd,
S.~T.~Johnson, T.~M.~Riseman, M.~P.~Nutley \\}
\address{School of Physics and Astronomy, University of Birmingham,
 Birmingham B15 2TT, UK.\\}

\date{\today}
\maketitle
\widetext

\begin{abstract}
We investigate the superconducting behaviour of Bi doped Pb. Pure lead
shows type-I behaviour entering an intermediate state in a magnetic
field. High dopings of Bi ($>$3$\%$) lead to type-II behaviour showing  a
mixed state, where the magnetic field penetrates the  superconductor in
the form of a flux lattice. At intermediate doping, the sample shows both
type-I and type-II behaviour depending on the temperature. This
arises because the Ginzburg-Landau parameter $\kappa$ passes through its
critical value of $1/\sqrt{2}$ with temperature.
\end{abstract}
\begin{multicols}{2}
\narrowtext
\section{INTRODUCTION}

When a superconductor is brought into a magnetic field, its
behaviour depends on the balance of its fundamental length scales.
The first of these, the penetration depth $\lambda$, describes the
length over which the applied field penetrates into the
superconductor. The second, the coherence length $\xi$, roughly
gives the spatial extent of the Cooper pair. Phenomenologically,
$\xi$ marks the length scale over which the macroscopic wave
function of the supercarriers varies. In the treatment of
superconductors by Ginzburg and Landau, this wavefunction takes
the role of the order parameter in the Landau theory of phase
transitions. The ratio of the two length scales, $\kappa = \lambda
/ \xi$, determines the response of the superconductor to an
applied magnetic field. In the case that $\kappa < 1/\sqrt{2}$,
the superconductor is said to be of type-I. Such a type-I
superconductor will expel a magnetic field completely, apart from
a layer at the surface of thickness $\lambda$. This is true as
long as the field is smaller than some critical field $B_c$,
associated with the de-pairing current. For applied fields higher
than $B_c$, superconductivity is lost and the field penetrates a
sample like a normal metal. The region of the thermodynamic B-T
phase diagram below $B_c$, in which the superconductor behaves as
an ideal diamagnet is also called the Meissner-state. If the
sample geometry is such that there are sizable demagnetising
effects, even a small applied field will exceed the critical field
the edges of the sample \cite{tinkham}. At these places the field
will be able to penetrate the superconductor. As a result, normal
conducting regions with a penetrating field of $B_c$, and
superconducting regions in the Meissner state (expelling the
field) will coexist in the sample. This state is called the
intermediate state. To minimise the surface energy of these
coexisting domains, the normal state-patches form an irregular
pattern, similar to that of Weiss-domains found in ferromagnets.

In addition to the Meissner-state, a type-II superconductor (with $\kappa
> 1/\sqrt{2}$) shows a second superconducting phase, in which it is
energetically more favourable
for the field to penetrate the sample in the form of quantised vortices
of magnetic flux. These flux lines mutually repell and form a regular
lattice, that is usually of hexagonal symmetry. This second
superconducting phase consists of the mixed state,
where the whole of the sample makes up a vortex lattice. The mixed
state extends up to a field $B_{c_2}$, at which the normalconducting cores
of the vortices start to overlap. For low fields
however, corresponding to the Meissner-state, and big demagnetising
fields, similar arguments to those given above apply. Again, the field
will exceed the critical field $B_{c_1}$, above which the sample enters
the mixed state, at some places. Thus in this situation, there will be a
coexistence
between regions of the sample in the Meissner-state, where the field is
expelled completely, with regions in the mixed state, where the field
penetrates the sample in the form of a vortex lattice with a mean field
corresponding to $B_{c_1}$.

In the present work, we study samples that have values of $\kappa$ close
to the boundary between type-I and type-II superconductors. By doping Pb
samples with small amounts of Bi, impurities are introduced to the
electronic structure. Such impurities lead to a reduction of the mean
free path of electrons and thus reduce the value of $\xi$. Therefore
small dopings of Bi lead to an
increase in $\kappa$. Hence PbBi alloys with sizable amounts ($>3\%$)
of Bi are type-II superconductors, whereas pure Pb is a typical example
of a type-I superconductor. With a suitable choice of Bi doping, the
coherence length and penetration depth are changed in such a way as to
permit a crossing of the temperature dependences of the critical fields
$B_c$ and $B_{c_2}$. These critical fields depend solely on the values of
the penetration depth and the coherence length and are thus strongly
affected by doping of the samples. The thermodynamic critical field can
be calculated to be:
\begin{equation}
B_c(T) = \frac{\Phi_0}{2\sqrt{2}\pi\lambda(T)\xi(T)},
\label{bc}
\end{equation}
where $\Phi_0$ = h/2e is the flux quantum. Its temperature dependence can
be calculated in the framework of the phenomenological two-fluid model to
be \cite{tilley}:
\begin{equation}
B_c (T) = B_c (0) (1- (T/T_c)^2).
\label{bct}
\end{equation}
Together with the temperature dependence of $\lambda$ that can be
independently calculated in the two fluid model ($\lambda^{-2} (T) =
\lambda^{-2} (0) (1-(T/T_c)^4)$, we obtain the temperature dependence of
$\xi$, which determines that of $B_{c_2}$, given by
\begin{equation}
B_{c_2} = \frac{\Phi_0}{2\pi\xi^2}.
\label{bc2}
\end{equation}
We thus obtain for the ratio of the two critical fields:
\begin{equation}
\frac{B_c}{B_{c_2}} (T) = \frac{1}{\sqrt{2}\kappa_0}(1+(T/T_c)^2) =
\frac{1}{\sqrt{2}\kappa_1}(T),
\end{equation}
where $\kappa_0 = \lambda(0) / \xi (0)$. From this we may then calculate
a crossover temperature $T^{I-II}$ between type-I
and type-II behaviour by equating $B_c$ and $B_{c_2}$. This means that at
temperatures above $T^{I-II}$, the thermodynamic phase transition will be at
$B_c$ and the superconductor will be
\begin{figure}
\input{epsf}
\epsfxsize 8.5cm
\centerline{\epsfbox{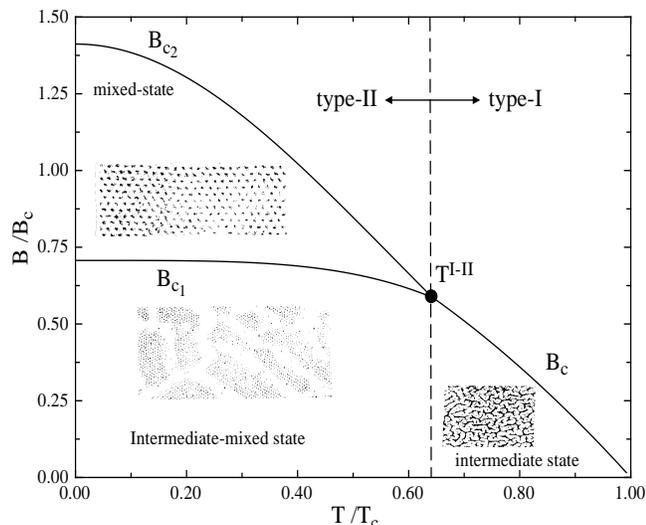}}
\caption{The thermodynamic phase diagram for a thin plate of a superconductor
with $\kappa_0$ =1. The different transition lines corresponding to type-I
and type-II superconducting behaviour meet in a multicritical point at a
temperature $T^{I-II}$, where the order of the transition changes from
first to second order. The phase lines are calculated from the two-fluid
model (see text).
}
\label{phasedia}
\end{figure}
\noindent of type-I. Below $T^{I-II}$ however, the
phase transition to the normal state happens at $B_{c_2}$ and thus the
superconductor is of type-II. From the above arguments, we obtain for
$T^{I-II}$
\begin{equation}
T^{I-II} = T_c (\sqrt{2}\kappa_0 -1)^{1/2},
\label{tstar}
\end{equation}
which only has a solution below $T_c$ for $1/\sqrt{2} < \kappa_0 < \sqrt{2}$,
putting a
severe limit on the observability of the effect. More exact values may be
determined from microscopic theories such as those of Helfand and
Werthamer,taking the electron mean-free paths into account \cite{HW}. The
results
however do not differ greatly and would only be available numerically,
thus obscuring the nature of the effects. In the B-T phase
diagram, this point at which the fundamental behaviour of the
superconductor changes may be interpreted as a multicritical point. An
overview
of this situation can be seen in Fig.~\ref{phasedia}, where we show the B-T
phase diagram of a superconductor with $\kappa_0$ =1. The different transition
lines $B_c$, $B_{c_2}$ and $B_{c_1}$ cross at the multicritical point at
$T^{I-II}$. Furthermore,
the phase transition at $B_c$ in a type-I superconductor is of first
order, whereas that into a type-II superconductor at $B_{c_2}$ is of
second order \cite{tilley}. The third phase line $B_{c_1}$, separating
the mixed state from the intermediate-mixed state in type-II superconductors
depends on the value of $\kappa$ and the nature of the interactions
between flux lines. For low $\kappa$ and attractive vortex interactions,
the transition is thought to be of first order \cite{auer}, otherwise it is
second order.
Thus we expect to observe a small region of coexistence between
type-I and type-II superconducting behaviour around $T^{I-II}$. We will
discuss this transition from type-I to type-II behaviour in more detail
below in the discussion of the experimental findings.

\section{EXPERIMENT}

We investigated four different samples of PbBi alloys, with different
doping levels of Bi. For a typical example of type-I behaviour, we used a
pure Pb sample and for a determination of the type-II properties we
investigated a sample with a doping of 5$\%$ Bi. The samples with
intermediate doping, 1.25 and 1.5$\%$, showed a transition from type-I to
type-II behaviour at different fields and temperatures. The samples all
were polycrystalline platelets of approximate dimensions 40x25x1 mm$^3$,
and an
ellipsoidal cross-section. The samples were characterised with a
vibrating sample magnetometer (VSM) and showed a superconducting transition,
$T_c$ of 7.2 K in zero applied field.

The muon spin rotation ($\mu$SR) experiments were carried out at the
MUSR instrument of the ISIS facility at the Rutherford Appleton
Laboratory (RAL), UK. The ISIS facility provides a pulsed beam of
positive muons at a frequency of 50 Hz with a pulse-width of $\simeq$ 70
ns. These muons are created in the decay of pions at rest and hence are
almost fully spin polarised, antiparallel to their momentum, and have a
kinetic energy of $\simeq$ 4 MeV.
The neutron scattering measurements were done at the Institut
Laue-Langevin (ILL), France, using the instrument D11.
ILL provides the world's most intense source of cold neutrons, where in
our investigations we have used neutrons of a wavelength of 1.9 nm. For
the very large structures in the intermediate mixed state, we needed to
\begin{figure}
\input{epsf}
\epsfxsize 8.5cm \centerline{\epsfbox{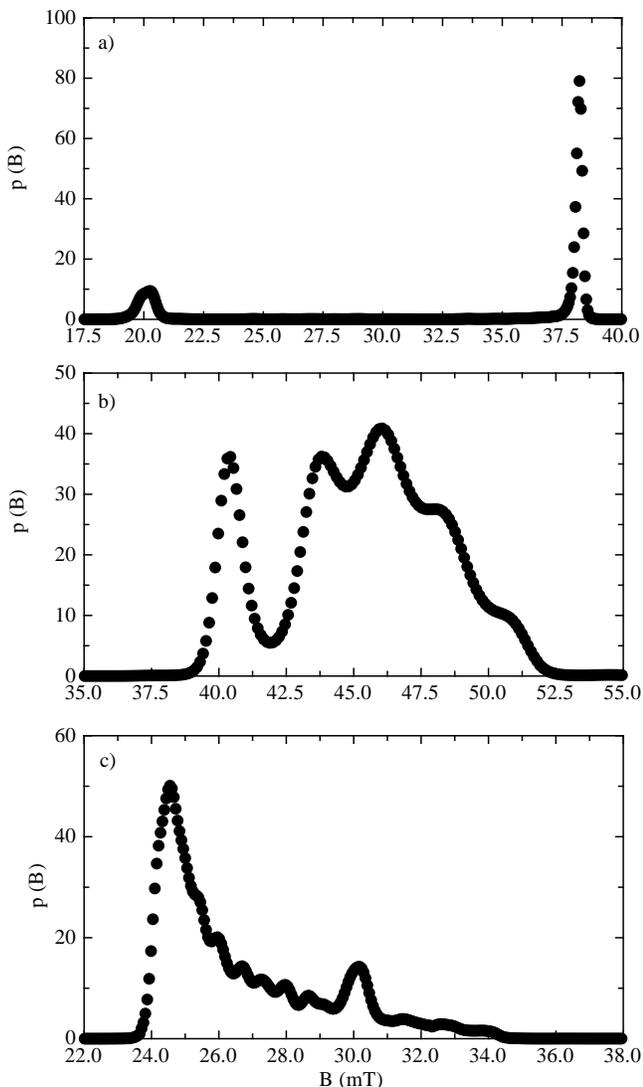}} \caption{a) The
local field distribution in the intermediate state. A background
of muons stopping in the cryostat shows an applied field of 20 mT,
while the muons in the sample precess at a frequency corresponding
to 38 mT. This corresponds to the critical field $B_c$ at the
temperature of 5.3 K, where the measurement was carried out. b)
The local field distribution in the intermediate mixed state. The
applied field can be seen to be 40 mT. In addition, there is a
wide distribution of local fields corresponding to a vortex
lattice at higher fields. The mean field of this flux lattice
corresponds to the critical field at the transition from type-I to
type-II behaviour $T^{I-II}$, if the overall field expulsion is
taken into account (see text). The measurement was carried out at
a temperature of 4.3 K. c) The field distribution for a vortex
lattice, in the mixed state. The field distribution has a very
characteristic shape, with a tail extending to fields higher than
the average. The measurement was taken in an applied field of 30
mT in the type-II sample containing 5 $\%$ Bi. } \label{muonp(B)}
\end{figure}
\noindent observe small-angle scattering, for which D11 with a distance
between the
sample and the detector of up to 40 m is ideal. In the investigations
described below, we have used a collimation length of 20.5 m and a
sample-detector distance of 20 m. No beam stop was used. The neutrons are
detected with a
position-sensitive detector of 0.8x0.8 m$^2$ area and pixel size 10 mm.
In both experiments, the field was applied at 45$^\circ$ to the shortest
direction of the sample, in order to have sizable demagnetisation fields.

In a transverse-field $\mu$SR experiment, positive muons come to rest in
the bulk of a sample, where the field is applied perpendicular to the
initial spin polarisation of the muons. Due to their magnetic moment, the
muons then perform a Larmor-precession with a frequency proportional to
the applied field. After a mean life-time of 2.2 $\mu$s, the muons decay
into a positron (and two neutrinos). Due to parity violation in the weak
decay of the muon, the decay-positron is preferentially emitted in the
direction of the muon spin. Therefore the number of positrons observed at
a fixed position, in the precession plane of the muons, will oscillate
with time reflecting the Larmor-precession of the muons, in addition to
an exponential decay, due to the radioactive decay of the muons. For a
distribution of fields over the sample, as is the case for a vortex
lattice, the oscillations due to the different fields superimpose giving
rise to a depolarisation of the muons. In the case of large field
gradients, the depolarisation rate may be fast enough to reduce the
initial polarisation outside of the experimental time window. In the mixed
state of a superconductor
with short penetration depth, this may strongly reduce the signal
observed by $\mu$SR. By studying the Fourier-transform
of the number of decay positrons with time, we may however still observe the
probability distribution of the internal fields, assuming a random
distribution of muons over the sample. In order to avoid spurious noise
in the Fourier-transform arising from the finite time window of
observation, we use a maximum-entropy algorithm in our analysis
\cite{rainford}. This
\begin{figure}
\input{epsf}
\epsfxsize 8.5cm
\centerline{\epsfbox{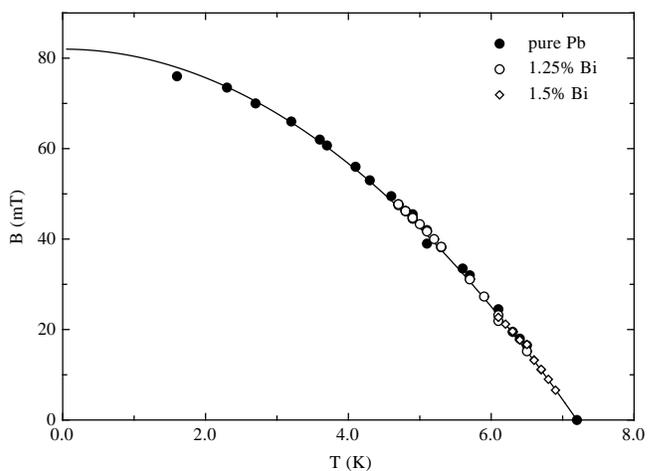}}
\caption{The thermodynamical critical field $B_c$, as determined from
$\mu$SR in the intermediate state. The measurements are for all
three samples exhibiting type-I behaviour. All curves were obtained with
several applied fields, where the different determinations overlap (see
text). It can also be seen that $B_c$ is the same for all samples in
spite of the change in $\kappa$.
}
\label{bcexp}
\end{figure}
\noindent also prevents statistical noise at long times, due to low count
rates, from
spreading over the whole frequency domain. This is mainly because our
algorithm
never actually Fourier-transforms the raw data, but rather solves the
inverse problem of calculating the frequency distribution and compares it
to the time-data. A more detailed discussion can be found in
Ref.~\cite{aegerter98}.

For the vortex lattice, in the mixed state, the field distribution has a very
specific shape, with a tail extending to fields higher than the average
\cite{sidorenko}. This is due to the high fields in the vortex cores.
In the intermediate state, we expect a
single-frequency signal at the critical field $B_c (T)$. The signal size
in the intermediate state is given by the area fraction of the normal
state domains. Neglecting the effects flux expulsion this is simply
proportional to B / B$_c$(T), leading to a decrease in signal with
decreasing temperature. Demagnetising effects due to the experimental
geometry of having the field at 45$^\circ$ to the plate however lead to a
much stronger reduction of signal size with temperature. This leads to a
dependence on the critical field $\propto 1/ (B-B_c)$, in reasonable
agreement with the experimental findings. This may also be
observed from the asymmetry in decay positrons upstream and downstream of
the beamline. This corresponds to muons not precessing in a field that
hence stopped in a Meissner domain in the sample. The temperature
dependence of this signal is complementary to that observed from the
normal state domains.

In a neutron scattering experiment, the neutrons are incident almost
parallel to the applied field. The alignment of the beam with respect to
the field was done by observing the diffraction pattern of  the vortex
lattice in a sample of Nb to an accuracy of 0.1$^\circ$ both vertically
and horizontally. Due to its magnetic moment, the neutron is scattered by
gradients of magnetic field \cite{nb}. Therefore we may observe the domain
pattern
in the intermediate state in terms of a correlation function. In general,
the scattered neutron intensity is given by \cite{squires}
\begin{equation}
I(Q) = \int R(Q) \frac{d\sigma(Q)}{d\Omega}dQ,
\end{equation}
where Q is the scattering vector, $R(Q)$ is a resolution
function, describing the distribution of neutrons around the
nominal scattering vector $Q$. Finally, $d\sigma / d
\Omega$ is given by \cite{squires}:
\begin{equation}
\frac{d\sigma(Q)}{d\Omega} \propto F(Q) S(Q),
\label{crossec}
\end{equation}
where $F(Q)$ is a form factor describing the scattering of a basic building
block of the structure and $S(Q)$ is a structure factor, describing the
large scale structure. The fourier transform of the structure factor
presents a measure of the `pair correlation' or distance distribution
function in the large scale structure. Thus eq.~\ref{crossec} corresponds
to a convolution of the correlation functions of
the two parts of the scattering.

In the intermediate state, this is a
similar problem to that encountered in the study of vesicles of
amphiphilic molecules that may build large scale structures. In contrast
to these amphiphilic molecules however, the microscopic structure of the
magnetic field domains is very simple. This microscopic structure is
described by the London equations describing the magnetic behaviour
of superconductors. Thus in the general expression for the scattered
intensity, the form factor will be given by
the well known London form factor. For the structure factor, we have
used a model of randomly oriented
chains with a radius of gyration R$_g$ in the analysis of the
measurements given below. This structure factor can be calculated to be
\cite{pedersen}
\begin{equation}
S_c(Q) = \frac{2(exp(-R_g^2Q^2)-1+R_g^2Q^2)}{R_g^4Q^4}.
\label{chains}
\end{equation}
In the limit of large $R_gQ$,
this form factor has the asymptotic dependence $S_c (Q) \propto Q^{-2}$,
corresponding to a random walk arrangement of the chains. From the
determination of $R_g$, we may then determine the contour length
\begin{figure}
\input{epsf}
\epsfxsize 8.5cm \centerline{\epsfbox{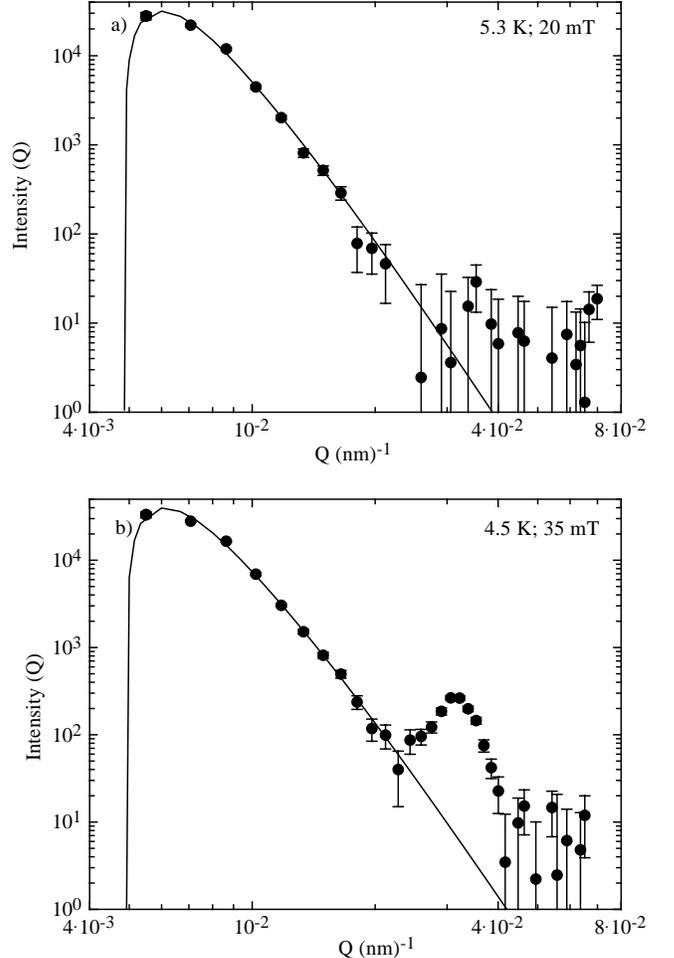}} \caption{a) The
scattered neutron intensity as a function of the scattering vector
$Q$. A diffraction pattern above $T_c$ has been subtracted. The
solid line presents a fit to a model using the structure factor of
eq.~\ref{chains}. Thus we determine the radius of gyration and the
interface area of the domain structure (see text). b) These
measurements were done in the intermediate-mixed state. In
addition to the structure factor from the domains in the
intermediate-state, there is a strong peak corresponding to Bragg
diffraction from the vortex lattice also present. The lattice
spacing corresponds to a field of 45 mT, the critical field at
$T^{I-II}$,the transition from type-I to type-II behaviour. The
solid line represents a fit to the same model as above (see text).
} \label{I(q)}
\end{figure}
\noindent of a chain, from $L = 6 R_g^2 /b$, where we assume the
statistical segment
length $b$ to be the penetration depth. This model will
break down below the length scale, where the magnetic domains become stiff.
On this scale, a model of randomly distributed, solid rods is more
appropriate. In this case, the structure factor has the asymptotic
dependence $S_r \propto Q^{-1}$. For the magnetic domains, in the
intermediate state, this length scale will be of the order of the
penetration depth ($\lambda \simeq 90 nm$). This makes it necessary to
observe very small scattering vectors, in order to resolve the domain
structure in the intermediate state. Due to the very small scattering
angles the observed scattering distribution is also influenced by the
width of the main beam. This may be taken into account by fitting a
Gaussian distribution to the main beam in the normal state. The intensity
of the main beam in the superconducting state is then reduced by those
neutrons scattered according to $S_c(Q)$, whereas we assume the width of
the main beam to be that in the normal state. From the radius of gyration
thus determined we may then obtain the contour length of a domain in the
intermediate state from
\begin{equation}
R_g^2 = \frac{L b}{6},
\label{length}
\end{equation}
where the Kuhn segment $b$ is of the order of the penetration depth
$\lambda$ and $L$ is the contour length. In Landau's treatment of the
intermediate state \cite{landau},the separation of the different domains
in an equilibrium situation depends only on the thickness of the plate,
the surface energy length $\delta$, usually associated with the coherence
length in type-I superconductors and a universal function f(b), where b
is the reduced field B/B$_c$. This function is calculated from the
different energy scales of the order parameter and the magnetic field and
is given by \cite{landau}
\begin{equation}
f(b) = \frac{1}{4\pi}[(1+b)^4 ln(1+b) +(1-b)^4 ln(1-b)
\end{equation}
\begin{displaymath}
- (1+b^2)^2 ln(1+b^2) -4b^2 ln(8b)].
\end{displaymath}
The separation of the different domains in the intermediate state is then
given by the expression
\begin{equation}
L =\sqrt{\frac{d \delta}{f(b)}},
\end{equation}
where $d$ is the thickness of the plate. In the case of interest here,
where $\kappa$ is close to the boundary of 1/$\sqrt{2}$, the surface
energy length becomes very small, such that the separation of the domains
may of similar order as the separation of vortices in the mixed state and
thus observable in neutron scattering experiments. This will be discussed
further in the context of the experimental findings.

In the mixed state, the situation is simpler, as the scattering now is
from a lattice. Therefore the structure factor will exhibit narrow peaks
when the scattering vectors equal a reciprocal lattice vector. The big
lattice spacing in the vortex lattice ($\sim$ 200 nm at 60 mT) leads
to very small angles of scattering ($\sim 1^\circ$). In the
intermediate-mixed state, there will be both structures, that of a vortex
lattice and the characteristic domain structure of the intermediate
state. The structure factor will thus be made up of a combination of both
Bragg and domain-scattering.

\section{RESULTS AND DISCUSSION}
A typical $\mu$SR measurement of the intermediate state in a type-I
superconductor is shown in Fig.~\ref{muonp(B)}a) for the case of the 1.25
$\%$ Bi sample at a temperature of 5.3 K. Due to a small background of
muons stopping in the cryostat, we observe the applied magnetic field
(20 mT). There is also a second
peak at the critical field $B_c$ corresponding to the temperature of 5.3
K. This field can be seen to be 38 mT in the figure. Thus we can map the
phase boundary $B_c (T)$ in the type-I state, by measuring the local
field in the sample at different temperatures. This is shown in
Fig.~\ref{bcexp} for the pure Pb sample, as well as the 1.25$\%$ Bi and
1.5$\%$ Bi samples in the type-I state. Due to the loss of signal with
increasing field gradient, the measurements were performed over an
overlapping range of applied fields. The observed $B_c (T)$ curve is
found to be independent of the applied field. Furthermore, the
temperature dependence of $B_c$ is in good agreement with the
\begin{figure}
\input{epsf}
\epsfxsize 8.5cm
\centerline{\epsfbox{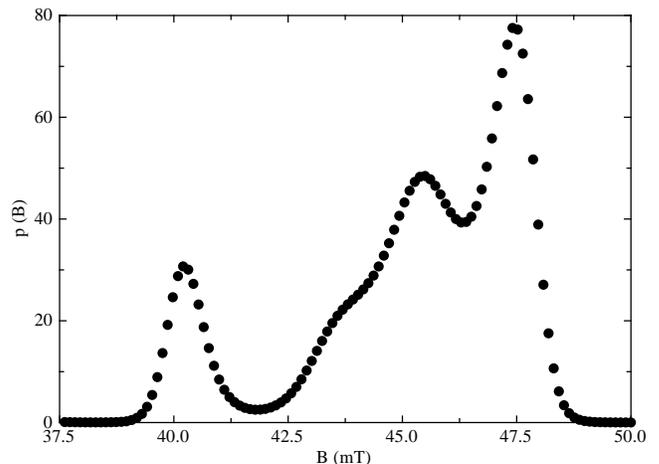}}
\caption{The distribution of local fields as observed by $\mu$SR in the
1.25$\%$ Bi sample at a temperature of 4.7 K. The applied field is 40 mT.
Moreover, one observes a sharp peak at a
field corresponding to the critical field at that temperature (see
Fig.~\ref{bcexp}). Finally one observes a very wide distribution of
fields below $B_c$. This corresponds to flux lines that have nucleated
out of the normalconducting regions in the intermediate state. Therefore
in this situation, one observes a coexistence of the intermediate and the
intermediate-mixed state, as would be expected at a multicritical
point (see text).
}
\label{coex}
\end{figure}
\noindent expectations from the two-fluid model, as can be seen from
the solid line in Fig.~\ref{bcexp}, representing a fit to eq.~\ref{bct}.
Moreover, it can be seen in the figure that for the doped samples the
value of $B_c$ remains unaltered. Considering the dependence of $B_c$ on
the values of the coherence length and the penetration depth
(eq.~\ref{bc}), we note that $\lambda'\xi' = \lambda \xi$, where
$\lambda'$ and $\xi'$ are the values of the penetration depth and
coherence length after doping. This agrees with the expectations of
microscopic theories on the influence of impurities on the
superconducting parameters in the temperature region of interest \cite{HW}.

A typical SANS measurement of the
intermediate state can be seen in Fig.~\ref{I(q)}a). This figure shows
the 1.25$\%$ Bi sample at 5.3 K, as Fig.~\ref{muonp(B)}a) does for the
$\mu$SR measurements. The intensity shown presents a subtraction of the
scattering observed at 5.3 K and that above $T_c$. The data presents a
radial average over the detector, giving the dependence of the scattered
intensity on the absolute value of Q. The solid line in Fig.~\ref{I(q)},
represents a fit to eq.~\ref{crossec}, where the structure factor was
assumed to be given by eq.~\ref{chains}. Furthermore we have taken into
account the effects arising from the very small scattering angles as
discussed above. The value for the
penetration depth used in calculating the London form factor was obtained
from the $\mu$SR measurements on the vortex lattice in this sample. From
this fit we determine the scattered intensity and the radius of gyration of
the domain structure. Here, the scattered intensity presents a measure of
the interface area between the domains and the radius of gyration roughly
gives the size of the domains. This shows the very complementary results
obtained from $\mu$SR and SANS.

Fig.~\ref{muonp(B)}b) shows the field probability distribution observed
in $\mu$SR in the intermediate-mixed state. The measurement was done in
the 1.25$\%$ Bi sample at a temperature of 4.3 K with an applied field of
40 mT. In addition to the background of muons stopping in the cryostat,
we observe a signal arising from a flux line lattice with a mean field
much higher than the applied. We will see below that the internal field
corresponds to $B_c (T^{I-II})$, the critical field at the temperature
where the superconducting behaviour changes. The same behaviour can be
seen in Fig.~\ref{I(q)}b), where the scattered neutron intensity is shown
at the same temperature, in an applied field of 35 mT. As in
Fig.\ref{I(q)}a), the scattered intensity
above $T_c$ has been subtracted and the London form factor has been taken
into account. Similar to Fig.~\ref{I(q)}a), we observe scattering from
the domain structure are low Q, that is similar to that observed in
Fig.~\ref{I(q)}a),
but in addition there is a peak in scattered intensity arising from the
vortex lattice in the intermediate-mixed state. The plane spacing of this
lattice, as obtained from the Q-value of the peak, corresponds to a mean
field of a triangular lattice of 42 mT, in good agreement with the results
from
$\mu$SR. The plane spacing of the vortex lattice is connected to the
field by the fact that each vortex line carries a quantum of flux. Thus
the plane-spacing is given by $d = (\sqrt{3}/2 \Phi_0 / B)^{1/2}$ for a
triangular lattice. From a comparison of Fig.~\ref{muonp(B)}b) with
Fig.~\ref{muonp(B)}c), where the field distribution of a vortex lattice
is shown, as obtained in the type-II sample containing 5 $\%$ Bi, it can
be seen that the field distribution in the intermediate-mixed state does
not perfectly match that of the mixed state. This may be understood from
a comparison of the lattice spacing with the separation of the
domains in the intermediate state. As will be shown below, they are of the
same order of magnitude ($L \simeq 10 \lambda$), which may lead to
imperfections in the vortex lattice resulting in a more symmetric lineshape.

The value of the mean internal field in the intermediate-mixed state can
be understood by considering the transition between type-I and type-II
superconducting behaviour. The penetration of the field in the form of a
vortex line arises from a competition between the surface energy gained by
expelling the magnetic field to the cost of an inclusion of
normalconducting material. The first will be roughly proportional to
$\lambda$, whereas the second will be proportional to $\xi$ (see
Ref.\cite{kopitzki}).
From this argument, it is seen that for a type-I superconductor, it is
energetically unfavourable to build up domain boundaries. A
rigorous calculation of these surface energies leads to the criterion of
$\kappa = 1/\sqrt{2}$ for the crossover from type-I to type-II behaviour
when these energies are equal. We now consider the case of the
intermediate-mixed state found in our samples at below $T^{I-II}$. On
cooling the sample, it is in the intermediate state with normalconducting
domains of mean field corresponding to $B_c (T)$. At the temperature
$T^{I-II}$, the ratio of the surface energies passes through one. Thus as
discussed above the formation of normalconducting-superconducting
boundaries becomes energetically favourable. This corresponds to a sign-change
of an effective surface tension of the magnetic field in the domains,
leading to the nucleation of vortices from the normalconducting domains,
due to flux quantisation in the superconductor. The domain structure however,
remains the same, which is why the mean field of the vortex lattice
corresponds to the value of $B_c (T^{I-II})$, due to flux conservation.

\begin{figure}
\input{epsf}
\epsfxsize 8.5cm
\centerline{\epsfbox{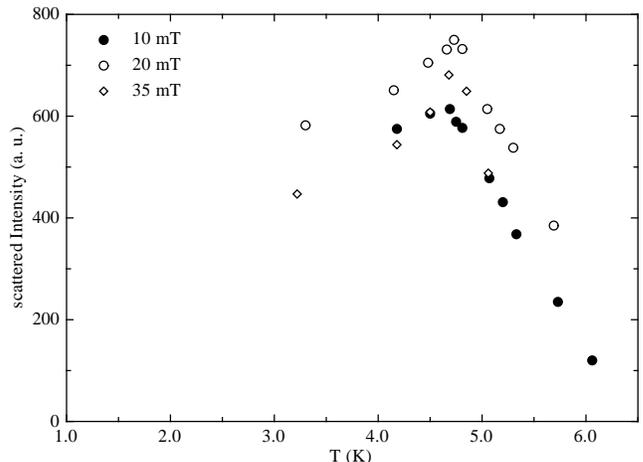}}
\caption{The scattered neutron intensity from the domain structure in the
intermediate and intermediate-mixed state as a function of temperature at
three different applied fields in the 1.25$\%$ Bi sample. The scattered
intensity was obtained by a fit using the model given by
eq.~\ref{chains}. This presents a measure of the interface area of
the domains, thus describing the ruggedness of the flux distribution in
the sample (see text). As would be expected the
scattered intensity shows a maximum around the temperature $T^{I-II}$,
where vortices are nucleating from the normal domains (see text).
}
\label{lowq}
\end{figure}

As the order of the superconducting transition changes between a type-I and
type-II superconductor in a magnetic field, the point in the B-T phase
diagram at $T^{I-II}$ may also be interpreted as a multicritical point. In
this scenario, we would expect to see a co-existence of the intermediate
and the intermediate-mixed state over a short temperature interval. That
this is in fact the case can be seen in Fig.~\ref{coex}, where we show
the local field distribution observed from $\mu$SR at a temperature of
4.7 K in the 1.25$\%$ Bi sample. As can be seen in the figure, in addition
to the sharp peaks at the applied field and $B_c$, there is a field
distribution at fields just below $B_c$. This coexistence of the
intermediate and the intermediate-mixed state is observed over a
temperature interval of 0.4 K. Within our temperature resolution of 0.1
K there is no hysteresis observable in the ratio between the two states.

From the onset of superconducting behaviour at higher fields, we
determine the value of $B_{c_2}$, which together with the value of $B_c$
determined from the Pb sample results in a determination of $\kappa_0$.
For the two samples showing a transition from type-I to type-II
behaviour, this results in $\kappa_0^{1.25}$ = 1.02(2) and $\kappa_0^{1.5}$
= 1.21(3). Using eq.~\ref{tstar}, we may then estimate the transition
temperature $T^{I-II}$, resulting in $T^{I-II}_{1.25}$ = 4.8(2) K and
$T^{I-II}_{1.5}$ = 6.1(1) K. These values are in excellent agreement with
those obtained directly in both $\mu$SR and SANS from the appearance of
the intermediate-mixed state. These directly determined values are
$T^{I-II}_{1.25}$ = 4.7(1) K and $T^{I-II}_{1.5}$ = 6.2(1) K. This also
indicates
that the phenomenological two-fluid model gives a fair description of the
superconducting behaviour of these PbBi alloys.

We now briefly return to the observation of the intermediate state with
SANS. As already noted above, from the structure factor we determined in
the intermediate state, we may obtain a measure of the interface area
from the total scattered intensity, as well as the radius of gyration of
the domain structure. Both these quantities depend upon temperature. At
high temperatures, close to $T_c$, the critical field is close to the
applied field, such that most of the sample area will be in the normal
state and the domain interface will be small. At low temperatures, the
critical field is much higher than the applied field and only a very
small area will be in the normal state. Thus again the domain interface
will be small. Therefore the scattered intensity will exhibit a maximum
with temperature, where the domain structure is most convolved. This can
be seen in Fig~\ref{lowq}, where we show the temperature dependence of
the scattered intensity in the 1.25$\%$ Bi sample at different applied
fields. As can be seen in the figure, the intensity shows a marked peak at the
transition temperature $T^{I-II}$ and then decreases slowly at low
temperatures. Such a sharp peak in the region of coexistence between the
intermediate and the intermediate-mixed state is possibly due to the strong
increase in surface area of the magnetic flux structures when the two states
coexist. At this point, the domain structure does not
change, but the normalconducting regions are nucleating vortex lines,
leading to both an enhancement in the surface area and the size of the
magnetic domain structures. The width of the coexistence is similar to that
found by $\mu$SR. The decrease in intensity is faster at higher applied
fields.
This may arise from the fact that these high fields are already sufficiently
close to  $B_{c_1}$, such that the superconductor would simply be in the mixed
state at low temperature. Hence there would be no scattering from
domains in that situation.

In the same way, we may investigate the size of the domains from the radius of
gyration observed in the intermediate state. Using the relationships
between the radius of gyration determined from the fits to the scattered
neutron intensities with the separation of the domains and Landau's
expression for this separation, we may also determine the universal
function f(b). With the known thickness of the sample of 1 mm, there are
two adjustable parameters left, whose order of magnitude is known. These
are the Kuhn length of the segments of the chains that is roughly given
by the penetration depth and the surface energy length that has to be
somewhat smaller than the coherence length. Using a segment length equal
to the penetration depth, the surface energy length $\delta$ = 10 nm has to
be chosen to obtain reasonable agreement with the expectations from Landau
\begin{figure}
\input{epsf}
\epsfxsize 8.5cm
\centerline{\epsfbox{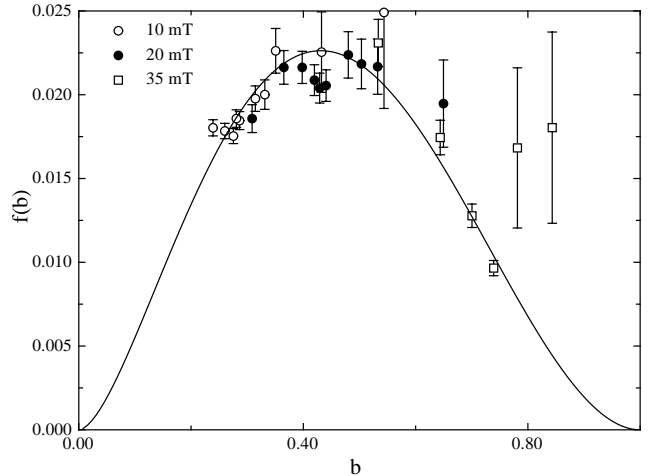}}
\caption{The function f(h) from Landau theory, as determined from the
radius of gyration of the normalconducting domains in the
intermediate and the intermediate-mixed state. The temperature
dependences at different fields allow us to determine f(h) over a big
range of reduced fields. In order to obtain reasonable agreement of the
size of the domains with the predictions of Landau theory, a surface
energy length of 10 nm, much smaller than the coherence length had to
be assumed. This is however not unreasonable, given the closeness of
$\kappa$ to 1/$\sqrt{2}$ (see text).
}
\label{rg}
\end{figure}
\noindent theory for f(b). These results can be seen in Fig.~\ref{rg} for
the 1.25 $\%$ sample. The
use of the temperature dependences at the different fields results in a
determination of f(b) over a big range of values. The value of $\delta$
used to obtain the agreement with Landau theory is an order of magnitude
smaller than the coherence length. This is however not unexpected, as the
value of $\kappa$ is very close to its boundary value of 1/$\sqrt{2}$ and
hence the surface energy length is expected to be small as discussed above.

\section{CONCLUSIONS}
We have measured the behaviour of type-I and type-II superconductors in a
magnetic field. It is shown that there is a transition from type-I to
type-II behaviour with temperature. This transition happens at a
temperature well defined in the framework of the two-fluid model and can be
shown to only arise for superconductors with a Ginzburg-Landau parameter
in the interval $1/\sqrt{2} < \kappa_0 < \sqrt{2}$ at low temperature.
This transition may be seen as a multicritical point in the B-T phase diagram
of superconductors, as the order of the superconducting transition is
changed from first to second order. Therefore we also observe a
coexistence of the different behaviours around the transition with both
SANS and $\mu$SR. We furthermore use the complementarities of SANS and
$\mu$SR to study the intermediate state in type-I superconductors. Using
SANS, we observe the domain structure via the interface area given by the
scattered intensity and the domain sizes via the radius of gyration,
where there is qualitative agreement with the treatment of the
intermediate state by Landau. The
local fields are investigated with $\mu$SR, where we can map the critical
field as a function of temperature.

\section{Acknowledgements}
We would like to thank the staff at ILL (Andreas Polzak) and ISIS
(Chris Scott, James Lord) for technical assistance, as well as the
Swiss National Science Foundation and the EPSRC of the UK for
financial support.

\bibliographystyle{prsty}

\end{multicols}

\end{document}